%% file: main.tex
\documentclass[conference]{IEEEtran}
\usepackage{cite}
\usepackage{amsmath,amssymb,amsfonts}
\usepackage{graphicx}
\usepackage{textcomp}
\usepackage{xcolor}
\usepackage{booktabs}
\usepackage{multirow}
\usepackage{array}
\usepackage{url}
\usepackage{tikz}
\usepackage{hyperref}
\hypersetup{
    colorlinks=true,
    linkcolor=blue,  
    citecolor=blue,  
    urlcolor=blue    
}
\newcommand{\etal}{\textit{et al. }{}}
\usepackage{enumitem}

\usetikzlibrary{positioning,arrows.meta,fit,backgrounds,shapes.geometric,calc}
\graphicspath{{figs/}}

\def\BibTeX{{\rm B\kern-.05em{\sc i\kern-.025em b}\kern-.08em
    T\kern-.1667em\lower.7ex\hbox{E}\kern-.125emX}}

\begin{document}

\title{How Far Can Lightweight LLMs Go in Intent-to-SLO Translation?}

\author{\IEEEauthorblockN{Sabbir Ahmed, Hazem Abbas, Salimur Choudhury}
\IEEEauthorblockA{\textit{School of Computing, Queen's University}, Ontario, Canada\\
\{sabbir.ahmed, hazem.abbas, s.choudhury\}@queensu.ca}
}

\maketitle

\begin{abstract}
Intent-based networking (IBN) promises QoS provisioning in which the network derives the configuration from operator-stated goals. 
Its first step, translating a natural-language intent into enforceable service-level objectives (SLOs) grounded in a live snapshot of the network state, increasingly relies on large language models (LLMs). 
These large models add cost and energy to the management loop and typically require multi-GPU, cloud-hosted infrastructure.
Can lightweight open LLMs take over this step dependably? 
We answer this question with a five-dimensional evaluation of ten LLMs with $<$10B parameters from five families, under four prompting setups, on 716 placement intents. We evaluate translation performance, intent validation, hallucination, energy impact, and robustness to diverse phrasing. 
The best configurations translate intents with a strict $F_1$ of 0.943 and reject up to 191 of 195 invalid intents.
With few-shot reasoning, the strongest models hallucinate unsupported QoS requirements in only about one response in a hundred, whereas the zero-shot setup struggles. A translated intent costs 0.044--0.148 Wh of GPU energy without reasoning, and added energy cost for reasoning pays off for the weaker models. Furthermore, we show that realistic rephrasing lowers the rejection of invalid intents in 15 of 16 configurations for the best-performing models, leaving robust intent validation as a key open challenge for reliable IBN. Our code and data are available: \href{https://github.com/ggck43/intent2slo}{intent2slo}.
\end{abstract}


\vspace{-0.2cm}
\section{Introduction}
\vspace{-0.2cm}
\label{sec:intro}

Edge--cloud service networks must meet fine-grained quality of service (QoS) and reliability requirements for every service they host. 
Intent-based networking (IBN) lets an operator state them as a goal, e.g., ``place the service on the edge, on a node with at least 4 CPU cores and the lowest memory utilization'', instead of configuring them by hand~\cite{rfc9315,3gpp28312}. 
Intent translation, the entry point of emerging management architectures like 3GPP intent-driven management~\cite{3gpp28312}, RAN application orchestration~\cite{habib2026tnse}, and edge service placement~\cite{ferrari2026manifest}, turns this goal into service-level objectives (SLOs) that an orchestrator can enforce, grounded in the current network state. 
A wrong, missing, or invented SLO can end in a wrong placement and a QoS violation. 
Modern large language models (LLMs) have made this translation practical~\cite{islam2026intentengine,manias2024semantic,habib2025intent,ferrari2026manifest,wang2024netconfeval}.


Most of these systems, however, rely on large cloud-hosted LLMs, which take the operator's monitoring data outside its domain and add cost to every intent. 
Lightweight open models have been used to translate intents into network device programs~\cite{angi2025llnet} and matched larger models in intent-based orchestration~\cite{phone2026ibn}. Existing studies evaluate mainly translation performance, often on small test sets or with large hosted models~\cite{ferrari2026manifest,islam2026intentengine}, and operator-backed benchmarks assess telecom knowledge and safety~\cite{gsma2025opentelco}, but not intent-to-SLO translation. 
A reliable intent translator must also refuse invalid intents, avoid inventing requirements and wasting energy, and cope with diverse phrasing. To our knowledge, no study has evaluated lightweight LLMs for intent-to-SLO translation on all of these dimensions together.


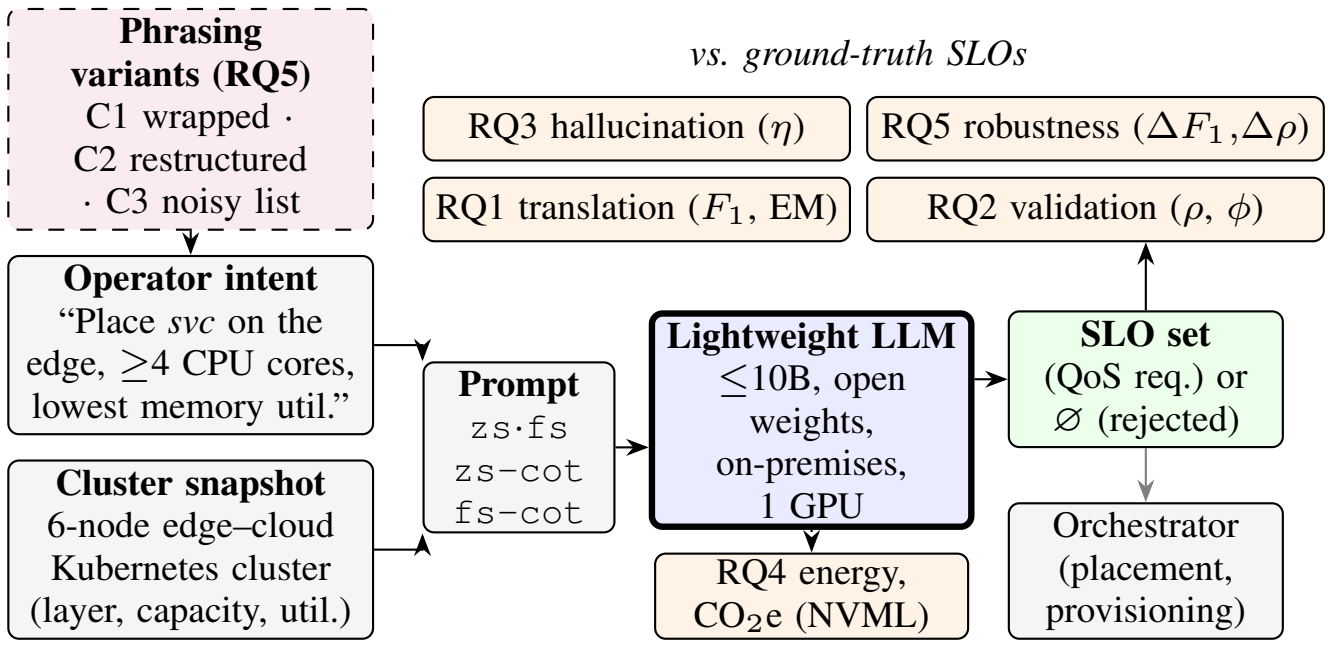
\begin{figure}[t]
\centering
\input{figs/fig1_single_d}
\caption{
Evaluation framework. A lightweight LLM translates an operator intent and the state snapshot of the edge–cloud cluster into SLOs, or rejects the intent. Prompts are zero-shot (\texttt{zs}) or few-shot (\texttt{fs}), with or without chain-of-thought (\texttt{cot}). The output is assessed for translation performance (RQ1), intent validation (RQ2), hallucination (RQ3), energy and CO$_2$e (RQ4), and robustness to operator phrasing (RQ5).
}
\label{fig:framework}
\vspace{-5mm}
\end{figure}

We fill this gap with a study of ten open lightweight LLMs from five families, four prompting setups, and 716 intents. 
We ask whether a lightweight model translates an intent into exactly the SLOs an orchestrator must enforce (RQ1), 
rejects ambiguous, conflicting, malformed, or unsupported intents (RQ2), 
and avoids inventing QoS requirements that neither the operator nor the network state supports (RQ3). 
We then quantify what each translation costs in energy and CO$_2$e and which deployment choices dominate that cost (RQ4), and test whether the translator holds when the same intents are rephrased in diverse ways (RQ5). \figureautorefname~\ref{fig:framework} summarizes the evaluation. 
Our answers form the contributions of this work:
\begin{itemize}
    \item \textbf{Translation performance (RQ1):} the best configurations reach a strict $F_1$ of 0.943,
    on par with the Intent Engine pipeline~\cite{islam2026intentengine}, and model family matters as much as size.
    \item \textbf{Reliable intent validation (RQ2):} the strongest models reject up to 98\% of invalid intents. 
    \item \textbf{Hallucination made measurable (RQ3):} we separate invented QoS requirements from ordinary mistakes; the best ones hallucinate in about one response in a hundred.
    \item \textbf{Energy implication (RQ4):} translating an intent costs 0.044--0.148\,Wh without reasoning; and reasoning multiplies this by 3.5--7.1$\times$, which pays off for weaker models. 
    \item \textbf{Robustness to operator phrasing (RQ5):} across three realistic rephrasing styles, both translation and validation performance degrade, with the rejection of invalid intents dropping up to 18.8 points. We release the 2{,}148 rephrased intents as a robustness benchmark.
  \end{itemize}

\vspace{-0.1cm}
\section{Related Work}
\vspace{-0.1cm}
\label{sec:related}

Manias \etal utilize semantic routing for LLM-assisted intent-based management of 5G core networks~\cite{manias2024semantic}, and Habib \etal extract the intent type and keywords for a hierarchical reinforcement-learning agent~\cite{habib2025intent}. 
Other works translate intents to generate deployable configurations, like programs for software-defined network devices with a small language model~\cite{angi2025llnet}, Linux traffic-shaping commands with 2.7B--8B open models~\cite{acharya2026intent2qos}, and Kubernetes resources for network-aware placement~\cite{ferrari2026manifest}. NetConfEval benchmarks LLMs on turning requirements into formal configurations~\cite{wang2024netconfeval}. 
Closest to our setting, Intent Engine translates placement intents into snapshot-grounded SLOs and reports rejection and hallucination only for GPT-4.1 mini~\cite{islam2026intentengine}, and Raptis \etal validate LLM-generated intents with symbolic guardrails~\cite{raptis2026validation}.
Most other works judge a translator only by how well its output matches a reference, often on small test sets (e.g., 30 requests~\cite{ferrari2026manifest}, 100 synthetic intents~\cite{acharya2026intent2qos}). 
Whether the model refuses intents it cannot serve, invents requirements, or what energy each translation costs is rarely measured. 

Phone \etal compare a fine-tuned 1.1B model with larger LLMs in an agentic intent-based orchestrator~\cite{phone2026ibn}. Ahmed \etal reported that many errors of 7B open models on telecom question answering (QA) are format failures~\cite{ahmed2024linguistic}. Hammar \etal show that a fine-tuned 14B model can plan incident response better than frontier LLMs~\cite{hammar2026hallucination}. These studies target QA, troubleshooting, or orchestration, but do not address intent-to-SLO translation that an orchestrator can enforce.

Definition of hallucination is often application-defined. Hammar \etal call an action hallucinated if it does not shorten the expected recovery and bound how often this happens through abstention~\cite{hammar2026hallucination}, Manias \etal report it only anecdotally~\cite{manias2024semantic}, and Intent Engine counts incorrect or missing values and schema violations alongside unsupported constraints~\cite{islam2026intentengine}. In natural language generation, it is content that the source contradicts~\cite{ji2023survey}. To our knowledge, no prior work separates invented QoS requirements from ordinary mistakes, or studies how lightweight translators handle ambiguous intents. 

Inference energy has been benchmarked per generated token for LLM serving~\cite{chung2025mlenergy}, and LLNet uses a small language model to make intent translation more sustainable~\cite{angi2025llnet}. However, generic benchmarks do not cover a network task with long snapshots and short structured answers, nor relate energy to performance and validation. 
Robustness to rephrasing has been studied for intent-based policy generation under linguistic distribution shift~\cite{zhang2026netinspector}, but not for intent-to-SLO translation. 
We bring these strands together by evaluating lightweight translators on performance, validation, hallucination, energy, and robustness to phrasing in one framework, which is the evidence needed before entrusting QoS provisioning to them.

\section{Evaluation Framework}
\label{sec:framework}


\subsection{System Model and Task}
\vspace{-0.1cm}
\label{sec:task}

We consider a service-oriented edge--cloud system managed through intents. A monitoring pipeline exports a cluster snapshot $\mathcal{N}$ that gives, for each node, its layer (edge or cloud), and its current capacity and utilization metrics. An intent $e$ for a service $s$ is a natural-language request.
The translation maps $(e, s, \mathcal{N})$ to a set of SLOs $\hat{S}$. Each SLO $\sigma$ is a QoS requirement with a type (placement, static capacity, dynamic utilization, or network resource), a metric, a comparator, a target value, and a unit~\cite{islam2026intentengine}. 
An \emph{explicit} SLO extracts a threshold from the intent (``at least 4 CPU cores''); an \emph{implicit} SLO comes from a superlative request (``the lowest memory utilization''), and its value must be read from the snapshot.
For a metric $m$ and a requested layer $\ell$, the value is
\begin{equation}
v^{*} = \operatorname*{opt}_{n \in \mathcal{N}_{\ell},\; x_m(n) \neq \varnothing} x_m(n), \qquad \operatorname{opt} \in \{\max, \min\},
\label{eq:grounding}
\end{equation}
where $\mathcal{N}_{\ell}$ is the set of nodes of layer $\ell$ (all nodes if the intent names no layer) and $x_m(n)$ is the value of $m$ at node $n$; nodes that report no value are skipped. Grounding ties the SLO to the current network state.

An intent that is ambiguous, self-contradictory, malformed, or asks for an unsupported metric must yield $\hat{S} = \varnothing$, as any SLO derived from it would steer provisioning toward requirements the operator never stated (\emph{intent validation}). The reference output of an intent is its \emph{ground-truth SLO} set $S$.

\subsection{Dataset}
\vspace{-0.1cm}
\label{sec:dataset}
We use the public dataset of Islam and Calheiros~\cite{islam2026intentengine} containing 716 placement intents on a 6-node edge--cloud Kubernetes cluster (four edge, two cloud nodes), all referring to one cluster snapshot. Of these, 521 are valid and 195 are invalid (65 ambiguous, 49 conflicting, 35 malformed, 46 unsupported). Each valid intent has a complexity level from 1 to 5, depending on the number and mix of its SLOs. The valid intents carry 1{,}169 ground-truth SLOs. 

\subsection{Prompt Design}
\vspace{-0.1cm}
\label{sec:prompts}
The prompts encode six requirements of the task, independent of the model: (i)~a \emph{fixed output schema} with the supported metrics;  (ii)~\emph{canonical units}, with explicit conversions; 
(iii)~\emph{grounding in snapshot}, reading superlative values from the snapshot; 
(iv)~\emph{no unstated requirements}, e.g., a placement only if the intent names a layer; (v)~\emph{intent validation}, returning an empty SLO list, 
for invalid intents, 
with instructions to limit false rejections; and (vi)~\emph{generic rules}, phrased without wording from the dataset's invalid intents.
The zero-shot prompt (\texttt{zs}) holds these instructions, the snapshot, and the intent. The few-shot prompt (\texttt{fs}) adds three invented examples, one per behavior: explicit thresholds with unit conversion, a conflicting intent, and implicit SLOs on a small invented snapshot; none resembles a dataset intent sample or shares a value with the real snapshot. The chain-of-thought variants (\texttt{zs-cot}, \texttt{fs-cot}) add an instruction to reason step by step before the final JSON object containing SLOs. 
Native reasoning (``thinking'') modes are disabled for all models.

\subsection{Metrics}
\vspace{-0.1cm}
\label{sec:metrics}

\textbf{Translation performance (RQ1).} As a misconfigured SLO is as harmful as a missing one, an SLO counts as correct only if its type, metric, comparator, qualifier (highest/lowest), value, and unit all equal those of a ground-truth SLO. Pooling true positives, false positives, and false negatives over all intents gives the strict micro-$F_1$, which weights every SLO equally. 
We also report the exact-match rate $\mathrm{EM} = \frac{1}{|V|}\sum_{i \in V} \mathbf{1}[\hat{S}_i = S_i]$ over the valid intents $V$, because the orchestrator can enforce an intent only when its whole SLO set is right.

\textbf{Intent validation (RQ2).} 
The rejection rate $\rho$ is the share of invalid intents answered with an empty SLO set $\hat{S}_i = \varnothing$. As over-eager rejection harms service availability, we also report the false-rejection rate $\phi$, the share of valid intents answered with $\hat{S}_i = \varnothing$.

\textbf{Hallucination (RQ3).} Not every deviation from the ground truth is a hallucination. A value off in its last digit, ``$<$'' for ``$\leq$'' or an empty value is a \emph{mistake}, as the model asserts nothing that the inputs contradict. 
An SLO is \emph{hallucinated} if it asserts a requirement or value that the intent text, its numbers, and the cluster snapshot do not support, e.g., 
a placement on a layer the operator never named, or a value that is not the requested metric's value in snapshot. 
The hallucination rate is the share of parseable responses $\mathcal{R}$ with at least one hallucinated SLO, $\eta = \frac{1}{|\mathcal{R}|}\sum_{r \in \mathcal{R}} \mathbf{1}[\exists\, \sigma \in \hat{S}_r : \sigma \text{ hallucinated}]$, since one unsupported requirement suffices to misconfigure a deployment.

The 28{,}640 responses from 10 models $\times$ 4 prompts $\times$ 716 intents contain 15{,}181 distinct (intent, SLO) pairs that do not match the ground truth, which is far beyond manual labeling. 
For valid intents, 
the hallucination rules (in the repository) are applied as deterministic rules against intent and snapshot. 
For invalid intents, we must infer ground-truth SLOs from free text.  
We used an LLM-assisted procedure with Claude Sonnet~5.5, 
a model not among those evaluated, 
which read each of the invalid intent texts under these rules and produced a per-intent specification (layers, metrics, directions, numbers), which was then applied automatically to every returned SLO. 
A human annotator checked 250 stratified labels (125 valid, 125 invalid), covering every rule at least twice and every model 25 times, and agreed with 93\% of cases.


\textbf{Energy and emissions (RQ4).} Energy efficiency is a design requirement of next-generation networks, and an always-on translation service adds to the energy budget of the infrastructure. We measure the GPU energy of each request with the NVIDIA management library (NVML) energy counter, the same source used by inference-energy benchmarks~\cite{chung2025mlenergy}. We report energy per intent (Wh), 
and operational emissions $\mathrm{CO_2e} = E \times \mathrm{PUE} \times \mathrm{CI}$, where $E$ is the energy, PUE is the power usage effectiveness of the data center and CI is the carbon intensity of the grid. Load levels, PUE, and grid scenarios are given in Section~\ref{sec:rq4}.


\textbf{Robustness (RQ5).} For each phrasing variant, we report the change $\Delta M = M_{\text{variant}} - M_{\text{original}}$ of a metric $M \in \{F_1,\rho\}$ from RQ1 and RQ2, so that a negative value is a loss caused by phrasing alone.

\subsection{Robustness to Operator Phrasing}
\vspace{-0.2cm}

\label{sec:robustset}
The dataset intents are short, one-line requests, whereas operators can write intents in quick notes, often with context, in lists, or in a hurry. 
We rewrote each intent in three compound variants (\tableautorefname~\ref{tab:robex}): \textbf{C1} (wrapped request) keeps the request close to the original or lightly reworded and wraps it with task-related opening and closing sentences; \textbf{C2} (restructured request) rewords the request, reorders its constraints, and adds one sentence on the service's role; 
\textbf{C3} (noisy list) turns it into a list with typically 3--5 typos and common shortcuts (e.g., ``pls'', ``mem''), followed by a short closing. 

\begin{table}[t]
\centering
\caption{
The three phrasing variants of one valid intent (RQ5).
}
\label{tab:robex}
\footnotesize
\input{tables/tab_robex}
\vspace{-5mm}
\end{table}

The rewriting rules preserve the meaning: numbers, units, service names, layer words, etc., are never changed by a typo.  Added sentences contain no numbers, metrics, or hints about validity; and every invalid intent keeps its defect (e.g., a malformed fragment stays broken). 
Each variant therefore keeps the ground-truth SLO set of its original intent.
The 716 intents yield 2{,}148 variants, which we release as a robustness benchmark. They were written one intent at a time with LLM assistance (Claude Sonnet~5.5), following the definitions (C1--C3) above, and automatic tests check that every variant keeps the numbers, units, service, layer, and superlative words of its original, and that added sentences are unique and digit-free. 
A human reviewer then checked all three variants of 100 randomly drawn intents (50 valid, 10 per level; 50 invalid, 12--13 per category) with 95\% agreement.  For each variant, the reviewer judged whether it keeps the ground-truth SLOs of its valid intents and, for an invalid intent, also its defect. 



\section{Results}
\vspace{-0.2cm}
\label{sec:results}
\subsection{RQ1: Translation Performance}
\vspace{-0.1cm}
\label{sec:rq1}
\begin{table}[t]
\centering
\caption{
Translation performance of the ten lightweight LLMs on the 521 valid intents; Mean is the $F_1$ averaged over the four setups. Best value per column in bold, second best underlined.}

\label{tab:rq1}
\setlength{\tabcolsep}{3pt}
\resizebox{\columnwidth}{!}{\input{tables/tab2_rq1}}
\vspace{-0.2cm}
\end{table}

Table~\ref{tab:rq1} reports the translation performance of all 40 configurations. $F_1$ ranges from 0.101 (Llama-3.2-3B, \texttt{zs-cot}) to 0.943 (Gemma-4-E4B, \texttt{fs-cot}), and the best exact-match rate is 0.879. 
Two configurations reach the $F_1$ reported for the full Intent Engine pipeline with GPT-4.1 mini (0.941)~\cite{islam2026intentengine}. 

The ranking by mean $F_1$ is led by Qwen3.5-9B (0.929) and Qwen3.5-4B (0.912). The 4B Qwen model has a higher mean than the 8B model of every other family. Within a family, gain from the larger model ranges from 0.016 (Qwen3.5) to 0.393 (Granite~4.1) in mean $F_1$. 
For an operator, choice of model family is therefore as important as the parameter budget. 

Our \texttt{fs-cot} gives the highest $F_1$ for 8 of the 10 models and the second highest for the other two.
The gain is largest for the models that are weak in \texttt{zs}: Ministral-3-3B rises from 0.408 to 0.847 and Gemma-4-E2B from 0.591 to 0.886. 

\begin{figure}[b]
\centering
\includegraphics[width=.93\columnwidth]{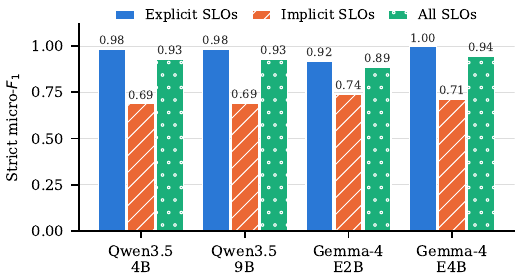}
\vspace{-0.1cm}
\caption{
Strict $F_1$ of the two best model families with \texttt{fs-cot} on explicit SLOs (thresholds stated in intent), implicit SLOs (values must be read from snapshot), and all SLOs. Errors lie mostly in implicit SLOs.
}
\label{fig:rq1}
\end{figure}

\figureautorefname~\ref{fig:rq1} separates explicit SLOs from implicit SLOs (whose value must be checked from snapshot). With \texttt{fs-cot}, the four Qwen and Gemma models reach an $F_1$ of 0.919--0.997 on explicit SLOs but only 0.688--0.739 on implicit ones. 



\subsection{RQ2: Intent Validation}
\label{sec:rq2}
\vspace{-0.1cm}
\begin{figure}[t]
\centering
\includegraphics[width=.9\columnwidth]{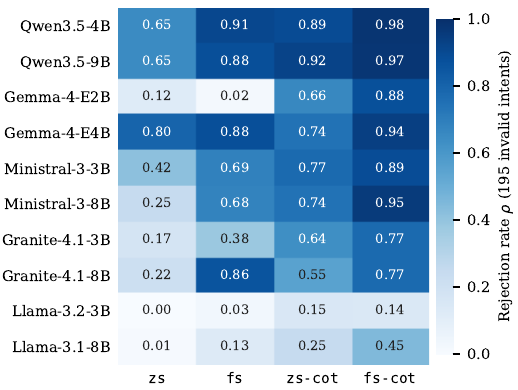}
\vspace{-2.5mm}
\caption{
Rejection rate $\rho$ of each model and setup on the 195 invalid intents (darker = more rejected). Examples and reasoning (\texttt{fs-cot}) reject most; whereas zero-shot setup and the Llama models reject least.
}
\label{fig:rq2}
\vspace{-0.3cm}
\end{figure}

An invalid intent that is translated into SLOs leads to provisioning that the operator never asked for. \figureautorefname~\ref{fig:rq2} shows how often each configuration rejects such intents. With \texttt{fs-cot}, the four strongest models reject almost all of them.
Zero-shot (\texttt{zs}) instructions are not enough, as $\rho$ ranges from 0.00 (Llama-3.2-3B) to 0.80 (Gemma-4-E4B), and only three models exceed 0.6. 
Examples and reasoning together (\texttt{fs-cot} against \texttt{zs}) raise $\rho$ for all ten models.

Among the 49 conflicting intents, contradictory bounds (e.g., a utilization below 40\% and above 80\%) and requests for both the highest and the lowest value of one metric are all rejected by Qwen3.5 and Ministral-3-8B with \texttt{fs-cot}. 
Intents that require the edge and the cloud at once are rejected by no model with \texttt{zs} (0 of 12) and by at most 10 of 12 with any setup (Ministral-3-8B, \texttt{fs-cot}). 
A layer conflict reads like a broad placement preference, so a model can satisfy it literally by writing both placements.

Over all configurations, 3{,}351 of the 7{,}800 answers to invalid intents contain SLOs. The most common is a single placement (22--50\% of them, depending on the category), which keeps the readable part of the request and silently drops the rest, and 363 answers to conflicting intents contain both halves of the contradiction.

Without reasoning, no model rejects any of the 521 valid intents. With reasoning, at most 6.9\% are rejected (Granite-4.1-3B, \texttt{fs-cot}). 
Qwen3.5-9B and Ministral-3-8B reject none of the valid intents. 
Reasoning thus makes most weaker models over-cautious on valid intents rather than more discerning.


\subsection{RQ3: Hallucination}
\vspace{-0.2cm}
\label{sec:rq3}

\begin{figure}[b]
\centering
\includegraphics[width=.9\columnwidth]{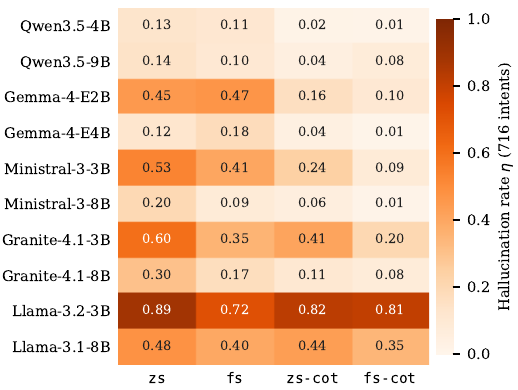}
\vspace{-2.5mm}
\caption{
Hallucination rate $\eta$ of each model and setup over all 716 intents (lighter = more trustworthy). Examples and reasoning (\texttt{fs-cot}) lower it for almost every model, down to about 1\% in the best configurations.
}
\label{fig:rq3}
\vspace{-0.2cm}
\end{figure}


\begin{table*}[t]
\centering
\caption{Hallucinated SLOs of both kinds: an unsupported value (grounded from the wrong metric) and an unsupported requirement (an invented placement); svc = service}
\label{tab:rq3}
\footnotesize
\input{tables/tab3_rq3}
\vspace{-0.2cm}
\end{table*}


\figureautorefname~\ref{fig:rq3} shows the hallucination rate per configuration. 
Reasoning lowers the rate for all ten models with zero-shot prompts and for nine of ten with few-shot prompts, and examples alone lower it for eight.
The best configurations hallucinate in about one response in a hundred: Gemma-4-E4B (0.008) and Qwen3.5-4B (0.013) 
with \texttt{fs-cot}. 

Pooled over its two models and four setups, $\eta$ per family ranges from 0.080 (Qwen3.5) through 0.191 (Gemma~4), 0.203 (Ministral~3) and 0.279 (Granite~4.1) to 0.610 (Llama~3). 
The larger model hallucinates less in four of the five families; Qwen3.5 is the exception, where the 9B model is slightly worse than the 4B model (0.093 against 0.068). 


On valid intents, eight of the ten models more often write an unsupported value (from the wrong node, metric, or unit conversion) than an unsupported requirement. On invalid intents, they fill in what the text leaves open, mostly a placement layer or a metric. \tableautorefname~\ref{tab:rq3} shows two examples where models hallucinated the requirements.



\subsection{RQ4: Energy and Emissions}
\vspace{-0.1cm}
\label{sec:rq4}

\begin{figure}[t]
\centering
\includegraphics[width=.93\columnwidth]{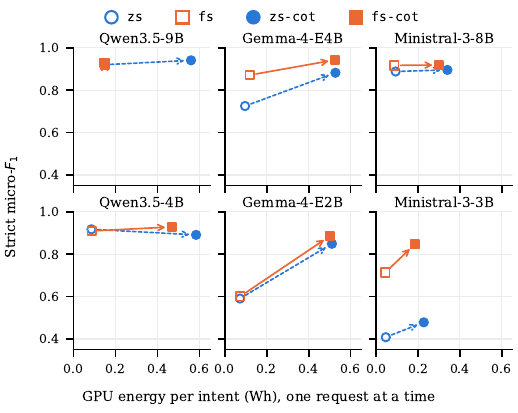}
\vspace{-3mm}
\caption{
$F_1$ against GPU energy per valid intent (one request at a time), one panel per model (columns: families; larger model on top), one point per setup. Arrows add reasoning (dashed: \texttt{zs}$\rightarrow$\texttt{zs-cot}; solid: \texttt{fs}$\rightarrow$\texttt{fs-cot}). Steep arrows are reasoning worth its energy; flat ones are not.
}
\label{fig:rq4}
\vspace{-0.3cm}
\end{figure}

An intent-translation service runs beside the infrastructure it manages, so its energy adds to the network's energy budget. 
We measured it for the three most accurate families in RQ1. 
Each model ran with vLLM on its own H100 NVL GPU with the prompts of RQ1, and no other process on it. 
Energy was read from the NVML counter~\cite{chung2025mlenergy}. 
We measured 11{,}808 requests at three load levels: one request at a time (as in a lightly used service), 16 and 64 concurrent requests (where batching raises efficiency at the cost of latency). 
No intent was repeated, so no complete prompt was served from the cache. 
Energies for all 521 valid intents are estimated from a per-model linear fit on generated tokens ($R^2 \geq 0.996$).
Served one request at a time, an intent costs 0.044--0.148\,Wh without reasoning and 0.18--0.58\,Wh with it (\figureautorefname~\ref{fig:rq4}). 
Reasoning multiplies the energy per intent by 3.5--7.1$\times$. This buys 26--28 $F_1$ points for Gemma-4-E2B, 13 for Ministral-3-3B (\texttt{fs}), and at most 2 points for Qwen3.5 and Ministral-3-8B. 
Qwen3.5-9B often answers the \texttt{fs-cot} prompt directly, without writing out its reasoning, so its \texttt{fs-cot} costs are similar to \texttt{fs}. Its \texttt{zs-cot} answers do contain reasoning.

Serving 64 requests at once instead of one at a time lowers the energy per intent by a factor of 3.2--20.2 (median 6.1). This happens most for the reasoning setups, whose long generations benefit most from batching.
The price is latency, as the 95th-percentile response time rises from 0.8--3.2\,s to 10--27.5\,s without reasoning and from 2.1--20.4\,s to 24.5--71.2\,s with it. So an operator can batch delay-tolerant intents (e.g., planned deployments) and serve urgent ones individually. 
At realistic rates, however, the GPU's idle power dominates the energy. For a dedicated GPU (idle draw 101--108 W) loaded to at most 50\% of its throughput under a 30 s latency bound, idle draw is 81--97\% (median 93\%) of the 2.5--3.2 kWh it consumes per day at 10{,}000 intents.
With an industry-average PUE of 1.56~\cite{uptime2024survey} and the 2025 world grid intensity of 0.458\,kg\,CO$_2$e/kWh~\cite{ember2026yearly, patterson2021carbon}, this is 0.18--0.67\,g\,CO$_2$e per intent. 
The data center's location changes it about 16-fold, from 0.016--0.061\,g in France to 0.26--0.98\,g in India.

\subsection{RQ5: Robustness to Operator Phrasing}
\vspace{-0.1cm}
\label{sec:rq5}

 
A reliable translator must hold for phrasings it was not evaluated with. We fed the 716 intents and their 2{,}148 variants (\sectionautorefname~\ref{sec:robustset}) with all four prompts to the two most accurate families, Qwen3.5 and Gemma~4, since phrasing losses are hard to separate from baseline errors in weaker models. 
For each configuration, we compare $F_1$ on the valid variants and $\rho$ on the invalid ones with the original intents.


\vspace{-0.1cm}
\begin{figure}[t]
\centering
\includegraphics[width=.95\columnwidth]{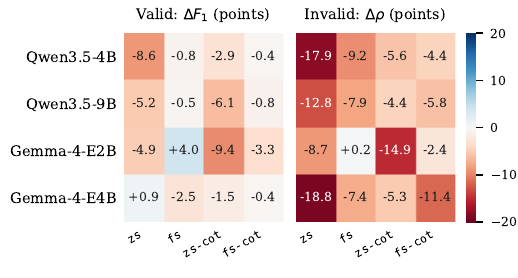}
\vspace{-5mm}
\caption{
Effect of operator phrasing per model and prompting setup. Change, in percentage, between the variants C1--C3 (pooled) and the original intents, for strict $F_1$ on valid intents (left) and  rejection rate on invalid intents (right). 
}
\label{fig:rq5}
\vspace{-4mm}
\end{figure}

\figureautorefname~\ref{fig:rq5} shows that rephrasing lowers the rejection rate in 15 of the 16 configurations by up to 18.8 points (Gemma-4-E4B, \texttt{zs}). $F_1$ drops in 14, by up to 9.4 points (Gemma-4-E2B, \texttt{zs-cot}). 
Zero-shot is the most fragile setup, losing 8.7--18.8 points of rejection. Few-shot examples reduce the losses, so that with \texttt{fs} or \texttt{fs-cot}, Qwen3.5 and Gemma-4-E4B lose 4.4--11.4 points of rejection and at most 2.5 points of $F_1$. 
The only configuration that gains on both metrics, Gemma-4-E2B with \texttt{fs} ($F_1$ +4.0, $\rho$ +0.2), does so from a weak baseline ($F_1$ 0.6, $\rho = 0.015$), not through robustness. Rephrasing thus costs some performance, but its more serious effect is that models accept invalid requests which they rejected in the short form.

\begin{figure}[b]
\centering
\includegraphics[width=.92\columnwidth]{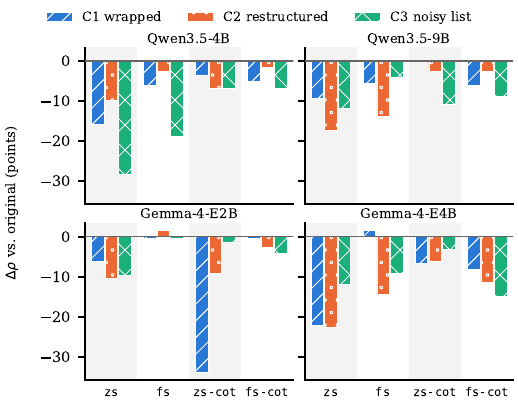}
\vspace{-3mm}
\caption{Change in rejection rate of invalid intents for each phrasing variant. 
}
\label{fig:rq5var}
\vspace{-0.4cm}
\end{figure}

\figureautorefname~\ref{fig:rq5var} shows the impact of the three variants separately. 
Restructuring (C2) and the noisy list (C3) hide contradictions, so the rejection of conflicting intents falls. Wrapping (C1) instead makes broken fragments read like normal requests, lowering the rejection of malformed intents.

\vspace{-0.1cm}
\section{Conclusion}
\vspace{-0.1cm}
\label{sec:conclusion}
We asked how far lightweight LLMs can go in intent-to-SLO translation. Our findings show that with few-shot reasoning, models of 4--9B parameters can reach a strict $F_1$ up to 0.943, stop up to 98\% of invalid intents, and rarely invent QoS requirements, at a fraction of a watt-hour per intent. 
Their limits are implicit values, intent validation that weakens under natural phrasing, and energy dominated by idle time. 
In practice, few-shot prompting should be the default, reasoning used for models where it adds performance, and the translator paired with a deterministic check of implicit values and SLO consistency. 
Our evaluation uses a single cluster snapshot, and energy was measured on an on-premises GPU. On edge accelerators, the absolute values will differ, while the relative effects of reasoning and batching are expected to hold.
These findings open research directions for AI-native network management, such as open benchmarks with realistic operator phrasing to harden intent validation, and energy-aware placement and scheduling of the translator on shared accelerators. Extending the evaluation to intent assurance in 3GPP intent-driven management and to multi-turn clarification of ambiguous intents will show how far lightweight models can carry closed-loop QoS management beyond translation.
\vspace{-2mm}

\bibliographystyle{IEEEtran}
\bibliography{refs}

\end{document}

%% file: figs/fig1_single_d.tex
\begin{tikzpicture}[
  font=\scriptsize,
  >={Stealth[length=1.5mm]},
  box/.style={draw, rounded corners=2pt, align=center, inner sep=1.6pt, minimum height=6mm},
  inp/.style={box, fill=gray!8},
  core/.style={box, fill=blue!8, very thick},
  outp/.style={box, fill=green!8},
  ev/.style={box, fill=orange!10, text width=27.5mm, minimum height=4.2mm, inner sep=1pt},
  var/.style={box, fill=purple!8, dashed},
  node distance=1.6mm and 2.2mm
]
\node[var, text width=23mm] (rob) {\textbf{Phrasing variants (RQ5)}\\C1 wrapped $\cdot$ C2 restructured $\cdot$ C3 noisy list};
\node[inp, text width=23mm, below=of rob] (intent) {\textbf{Operator intent}\\``Place \textit{svc} on the edge, $\geq$4 CPU cores, lowest memory util.''};
\node[inp, text width=23mm, below=of intent] (snap) {\textbf{Cluster snapshot}\\6-node edge--cloud Kubernetes cluster (layer, capacity, util.)};
\draw[->, dashed] (rob) -- (intent);
\node[ev, text width=20mm, anchor=south] (e4) at ($(snap.south east)+({3.2mm+11.5mm+3.6pt+2.2mm+0.5*(20mm+4.4pt)},0)$) {RQ4 energy, CO$_2$e (NVML)};
\node[core, text width=20mm, above=1.5mm of e4] (llm) {\textbf{Lightweight LLM}\\$\leq$10B, open weights,\\on-premises, 1 GPU};
\coordinate (mid) at ($(intent.east)!0.5!(snap.east)$);
\node[box, fill=gray!8, text width=11.5mm, anchor=east] (prompt) at ($(llm.west |- mid)+(-2.2mm,0)$) {\textbf{Prompt}\\\texttt{zs}$\cdot$\texttt{fs}\\\texttt{zs-cot}\\\texttt{fs-cot}};
\draw[->, dotted] (llm) -- (e4);
\draw[->] (intent.east) -- (prompt.west |- intent.east) -- (prompt.north west);
\draw[->] (snap.east) -- (prompt.west |- snap.east) -- (prompt.south west);
\draw[->] (prompt.east) -- (prompt.east -| llm.west);
\node[outp, text width=17mm, anchor=north west] (out) at ($(llm.north east)+(2.2mm,0mm)$) {\textbf{SLO set}\\(QoS req.) or\\$\varnothing$ (rejected)};
\draw[->] (llm.east |- out.west) -- (out.west);
\node[box, fill=gray!8, text width=17mm, anchor=south] (orch) at (out.south |- snap.south) {Orchestrator (placement, provisioning)};
\draw[->, gray] (out) -- (orch);
\node[ev, text width=27.5mm, anchor=south west] (e1) at ($(prompt.west |- out.north)+(0,4.5mm)$) {RQ1 translation ($F_1$, EM)};
\node[ev, text width=29.5mm, right=1mm of e1] (e2) {RQ2 validation ($\rho$, $\phi$)};
\node[ev, text width=27.5mm, above=1mm of e1] (e3) {RQ3 hallucination ($\eta$)};
\node[ev, text width=29.5mm, above=1mm of e2] (e5) {\mbox{RQ5 robustness ($\Delta F_1,\!\Delta\rho$)}};
\draw[->] (out.north) -- (out.north |- e2.south);
\node[font=\scriptsize\itshape, anchor=south] at ($(e3.north east)!0.5!(e5.north west)+(0,0.3mm)$) {vs.\ ground-truth SLOs};
\end{tikzpicture}

%% file: tables/tab_robex.tex
\resizebox{0.9\columnwidth}{!}{%
  \setlength{\tabcolsep}{3pt}\begin{tabular}{@{}>{\raggedright\arraybackslash}p{0.07\columnwidth}>{\raggedright\arraybackslash}p{\dimexpr0.93\columnwidth-6pt\relax}@{}}
  \toprule
  \textbf{Orig.} & For svc, please run it on the edge side, make sure the target node has at least 2 GB of memory, and pick the node with the highest storage capacity. \\
  \cmidrule(l){2-2}
  \textbf{C1} & We are retiring a couple of old machines at the end of the month, and the recommender is waiting for a new home. Please run svc on the edge side with a node that offers at least 2 GB of memory, and choose the one with the highest storage capacity. I'll pass the result to the migration team tomorrow, so thank you for a quick reply. \\
  \cmidrule(l){2-2}
  \textbf{C2} & The recommender has moved between hosts twice since spring and the team wants it settled. Choose the node with the highest storage capacity for svc, make sure it has 2 GB of memory or more, and keep it on the edge side. \\
  \cmidrule(l){2-2}
  \textbf{C3} & Service: svc\newline
  - Layr: edge side pls\newline
  - Memory: node need at least 2 GB of mem\newline
  - Storage: pick the node what have the highest storge capacity\newline
  Migration is soon so asap if you can. Thx, i really apreciate it. \\
  \bottomrule
  \end{tabular}%
}

%% file: tables/tab2_rq1.tex
\begin{tabular}{llccccccccc}
\toprule
 & & \multicolumn{2}{c}{\texttt{zs}} & \multicolumn{2}{c}{\texttt{fs}} & \multicolumn{2}{c}{\texttt{zs-cot}} & \multicolumn{2}{c}{\texttt{fs-cot}} & Mean \\
\cmidrule(lr){3-4}\cmidrule(lr){5-6}\cmidrule(lr){7-8}\cmidrule(lr){9-10}
Family & Size & $F_1$ & EM & $F_1$ & EM & $F_1$ & EM & $F_1$ & EM & $F_1$ \\
\midrule
Qwen3.5 & 4B & \underline{0.918} & \underline{0.820} & 0.910 & 0.774 & 0.892 & 0.772 & 0.929 & \underline{0.845} & \underline{0.912} \\
 & 9B & \textbf{0.921} & \textbf{0.835} & \textbf{0.922} & \textbf{0.802} & \textbf{0.942} & \textbf{0.879} & \underline{0.930} & 0.823 & \textbf{0.929} \\
\addlinespace[2pt]
Gemma 4 & E2B & 0.591 & 0.397 & 0.602 & 0.399 & 0.850 & 0.695 & 0.886 & 0.747 & 0.732 \\
 & E4B & 0.726 & 0.403 & 0.872 & 0.733 & 0.883 & 0.754 & \textbf{0.943} & \textbf{0.879} & 0.856 \\
\addlinespace[2pt]
Ministral 3 & 3B & 0.408 & 0.159 & 0.714 & 0.493 & 0.479 & 0.236 & 0.847 & 0.702 & 0.612 \\
 & 8B & 0.889 & 0.779 & \underline{0.917} & \underline{0.795} & \underline{0.896} & \underline{0.779} & 0.920 & 0.825 & 0.905 \\
\addlinespace[2pt]
Granite 4.1 & 3B & 0.248 & 0.188 & 0.651 & 0.359 & 0.269 & 0.150 & 0.522 & 0.253 & 0.423 \\
 & 8B & 0.764 & 0.587 & 0.855 & 0.672 & 0.776 & 0.601 & 0.867 & 0.716 & 0.816 \\
\addlinespace[2pt]
Llama 3 & 3B & 0.114 & 0.035 & 0.440 & 0.167 & 0.101 & 0.006 & 0.470 & 0.140 & 0.281 \\
 & 8B & 0.448 & 0.286 & 0.749 & 0.507 & 0.433 & 0.200 & 0.756 & 0.501 & 0.597 \\
\bottomrule
\end{tabular}

%% file: tables/tab3_rq3.tex
\resizebox{0.9\textwidth}{!}{%
\begin{tabular}{@{}>{\raggedright\arraybackslash}p{0.30\textwidth}>{\raggedright\arraybackslash}p{0.18\textwidth}>{\raggedright\arraybackslash}p{0.25\textwidth}>{\raggedright\arraybackslash}p{0.20\textwidth}@{}}
\toprule
Intent (abridged) & Ground-truth SLOs & Model (setup): hallucinated SLO & Why it is a hallucination \\
\midrule
``Deploy svc \ldots\ at least 50\,GB of storage capacity and pick the node with the lowest current memory utilization'' (valid) &
storage cap.\ $\geq$ 50\,GB; memory util.\ $\leq$ [lowest] 34.27\,\% &
Qwen3.5-9B (\texttt{zs}): memory util.\ $\leq$ [lowest] \textbf{18.0}\,\% &
18.0 is the \emph{storage} utilization of the correct node: right node, wrong metric \\
\addlinespace[2pt]
``I would like to place svc and keep CPU utilization below 40\,\%'' (valid) &
CPU util.\ $\leq$ 40\,\% &
Qwen3.5-9B (\texttt{fs-cot}): placement = \textbf{edge} &
a placement layer the intent does not name \\
\bottomrule
\end{tabular}
}